\documentclass[twocolumn,showpacs,preprintnumbers,prb,floatfix,amsmath]{revtex4}
\usepackage{graphicx}
\usepackage{times}
\usepackage{color}
\usepackage{dcolumn}
\newcommand{\mypath}[1]{./#1}

\begin{document}

\title{Insulator-to-Metal Transition Induced by Disorder \\
in a Model for Manganites}

\author{C. \c{S}en}
\author{G. Alvarez}
\author{E. Dagotto}
\affiliation{National High Magnetic Field Lab and Department of Physics,
Florida State University, Tallahassee, FL 32310}

\date{\today}

\begin{abstract}

The physics of manganites appears to be dominated by phase competition
among ferromagnetic metallic and charge-ordered antiferromagnetic
insulating states. Previous investigations (Burgy {\it et al.},
Phys. Rev. Lett. {\bf 87}, 277202 (2001))
have shown that quenched
disorder is important to smear the first-order transition between those
competing states, and induce nanoscale inhomogeneities that produce the
colossal magnetoresistance effect. Recent studies (Motome {\it et al.}
Phys. Rev. Lett. {\bf 91}, 167204 (2003))
have provided further evidence that disorder is important in the
manganite context, unveiling an unexpected
insulator-to-metal transition triggered
by disorder in a one-orbital model with cooperative phonons. In this
paper, a qualitative explanation for this effect is presented. 
It is argued that the transition occurs 
for disorder in the form of local random energies. Acting
over an insulating states made out of a  checkerboard arrangement of charge,
with ``effective'' site energies positive and negative,
this form of disorder can produce lattice sites with an effective energy
near zero, favorable for the transport of charge. This explanation is
based on Monte Carlo simulations and the study of simplified toy models,
measuring the density-of-states, cluster conductances using the
Landauer formalism, and other
observables. The applicability of these ideas to real manganites is
discussed.
\end{abstract}

\maketitle

\section{Introduction}

The study of manganites -- the materials with the colossal magnetoresistance --
continues attracting the attention of condensed matter experts.\cite{tokura,book}
These compounds are interesting for their potential technological applications as ``read
sensors'' in computers, as well as for the basic science challenges that
their unusual properties represent for our understanding 
of transition-metal oxides. It is widely believed that strong correlations are crucial
to understand these manganese oxides, either in the form of large Coulombic
interactions or large electron Jahn-Teller phonon couplings, or both
simultaneously. Some of their unusual properties include a remarkable
response to magnetic fields, with the DC resistivity changing 
by 10 or more orders of magnitude in some compounds upon the application 
of fields of order 1 Tesla. Another interesting property is the rich phase diagram
that these materials present, with competition of very different states
mainly involving ferromagnetic (FM) metallic and charge-ordered (CO) antiferromagnetic (AF)
phases. Moreover, in recent years a plethora of experiments
have unveiled a remarkable tendency to form inhomogeneities, which occur mainly
at the nanoscale but with some investigations reporting domains as large as submicrometer
in size.\cite{cheong} These spontaneous generation of nanoclusters was predicted
theoretically when phase separation tendencies were unveiled
in the first Monte Carlo simulations of models for
manganites in 1998 by Yunoki {\it et al.},\cite{yunoki}
then confirmed and refined experimentally in dozens of efforts.\cite{review}
The successful cross-fertilization between theory and experiment is remarkable 
in the area of Mn-oxides,
leading to considerable progress in recent years. These studies
may have consequences not only for manganites, but for a variety of
other compounds where inhomogeneities have been found in 
experiments, including the famous high-temperature
superconductors.\cite{davis}

Among the leading candidates to understand the CMR effect is the
`phase separation' scenario.\cite{book,review,burgy}
In this context, the state that is believed
to be abnormally susceptible to external magnetic fields is made out
of coexisting clusters of two competing phases. The clusters involving the FM state
have random orientations of their magnetic moments,
leading to an overall vanishing magnetization. However, the ``building blocks''
of the FM state (i.e. the FM nano clusters) are preformed and for this reason the state reacts
rapidly to magnetic fields, aligning the nanocluster moments in the
presence of small fields. Energetically it is
quite different to form a FM state from ``scratch'', as opposed
to having preformed FM islands that only need to reorient their magnetic moments
to generate global ferromagnetism. 
The competing state is also important in the proposed CMR state, since
it provides walls that prevent the alignment of the nanoclusters in
the absence of external fields.\cite{review,burgy}

In order to stabilize this type of CMR states, the role of quenched
disorder appears to be important.\cite{burgy} This disorder prevents the system
from being totally ferromagnetic or totally CO-AF, as it occurs in
the clean limit where a first-order phase transition separates the
states.\cite{review,book} Small amounts of quenched disorder locally favor one phase
over the other, leading to the inhomogeneous patterns found in simulations,
which are believed to correspond to those in experiments.
These predictions were beautifully confirmed by the group of
Tokura,\cite{tokura_recent} by means of a careful study of a
particular Mn-oxide that spontaneously forms a structure very close to the clean limit,
since all ions order in a regular pattern. This compounds presents a ``bicritical''
phase diagram, with a first-order FM-AF transition in excellent
agreement with Monte Carlo simulations.\cite{aliaga} When the material is rapidly
quenched in the growing process, such  that disorder is now incorporated in
the distribution of trivalent ions, then a behavior characteristic
of many other manganites is recovered. In particular, a large CMR
is found in the presence of disorder, result compatible with the
theoretical predictions.\cite{review,burgy}
Recent investigations\cite{burgy2} have shown that the addition of lattice elastic
effects (due to the cooperative nature of the Jahn-Teller distortions)
 makes the disorder strength needed to induce the
CMR phenomenon smaller in magnitude than without
elasticity, and avoids limitations in the critical dimension
that the naive use of Imry-Ma argumentations\cite{imry} would suggest.
The relevance of elastic effects has been remarked
extensively in recent literature as well.\cite{khomskii} Other power-law
decaying effects, such as unscreened Coulomb interactions, have also
been proposed as relevant in this context.\cite{yang} In addition,
investigations of a variety of models of percolation have shown
that ``correlated'' disorder -- to mimic elastic effects --
can even change the order of the transition rendering it first
order.\cite{first-order} It may occur that ``infinitesimal''
disorder is enough to
trigger and stabilize a phase separation
process that is intrinsic to systems with first-order phase transitions.
Nevertheless, in practice at least small amounts of disorder appear to be crucial
to produce the inhomogeneous state that theorists proposed as the key factor
to understand the CMR physics.

More recent investigations by Motome, Furukawa, and Nagaosa
\cite{motome} have provided important information for the
theoretical understanding of manganese oxides. Monte Carlo
simulations of the one-orbital double-exchange model including
cooperative phonons have revealed an interesting transition from
an insulator to a metal upon the introduction of quenched disorder 
into a CO state. The disorder is in the
form of on-site random energies. Once again, the
key role of disorder is unveiled by these simulations.
Independently, Aliaga {\it et al.}\cite{aliaga} arrived to
analogous conclusions studying a two-orbital model for manganites
including cooperative Jahn-Teller phonons. These authors noticed
that disorder drastically affects the CE charge-ordered state,
transforming it into a ``CE glass''. Overall, it is clear that the use of
cooperative phonons (for previous work see Refs.~\onlinecite{hotta,verges1}
and references therein), as opposed to the simpler case of
non-cooperative lattice distortions, is important to unveil interesting effects of
manganite models, that may be of relevance to experiments.

It is the purpose of this work to provide an intuitive understanding
of the results of Ref.~\onlinecite{motome}, mainly addressing 
the transition from
an insulator to a metal upon the introduction of disorder, a result
which seems counter-intuitive at first sight. Here, it is argued
 that the transition
occurs through an interesting percolative-like process due to the effect
of random on-site energies over a charge-ordered state. The sites
of the lattice that had an effective ``negative energy'' in the CO state
-- namely those that were populated with localized
electrons in the checkerboard arrangement --
may acquire a nearly zero total energy due to the influence 
of that type of disorder. The ``nearly zero energy''
sites can contribute to form a metallic state which is highly inhomogeneous
in instantaneous Monte Carlo snapshots, although it slowly 
evolves with time (dynamical) as the Monte Carlo simulation progresses. 
It is concluded that the case
found in Refs.~\onlinecite{aliaga,motome} provides
another very interesting example where a combination of explicit
quenched disorder and charge-ordering tendencies leads to nontrivial
physics, including insulator-metal transitions.

The organization of the paper is as follows. In Section II, the model
with cooperative phonons and disorder is discussed. The main results
of the Monte Carlo analysis are in Section III, including the phase
diagram, density-of-states with and without disorder, and calculation
of conductances. A simplified model that captures the essence of the
main model is introduced in Section IV. Both models
behave very similarly. Our proposed explanation of the insulator-metal
transition is presented in Section V, where the inhomogeneous
nature of the metallic state is described. Simulations with
random hoppings and by freezing the localized spins are also reported
in this section. Conclusions are presented in Section VI.

\section{Model}

The Hamiltonian used in this study contains Mn $e_{\rm g}$
itinerant electrons coupled to the $t_{\rm 2g}$ Mn core-spins, and
also to phonons according to:\cite{verges1}
\begin{eqnarray}
H=-t\sum_{\langle {\bf i},{\bf j} \rangle, \sigma}(c_{{\bf i}\sigma}^\dag c_{{\bf j}\sigma} + \mbox{h.c.})
- 2 J_{\rm H}\sum_{\mathbf{i}}\mathbf{s_{i}}\cdot \mathbf{S_i} \nonumber \\
- \lambda t \sum_{{\bf i}, \alpha, \sigma}(u_{{\bf i},-\alpha}-u_{{\bf i},\alpha})c_{{\bf i}\sigma}^\dag
c_{i\sigma}.
\end{eqnarray}
Here, the first term represents the kinetic energy of the carriers
hopping between Mn atoms
on a $d$-dimensional cubic lattice, with $t$ the hopping amplitude for the process.
$\langle {\bf i,j} \rangle$ denotes nearest-neighbor pairs of sites, and
the rest of the notation is standard.
The second term is the Hund interaction between itinerant
${\bf s}_{\bf i}$ and localized $\bf {S_i}$ spins,
with $J_{\rm H}$ the Hund coupling constant.
The third term describes the energy corresponding to the lattice-carrier interaction,
with $\lambda$ denoting the electron-phonon
coupling. Appropriate redefinitions of variables allow us to have the energy unit $t$
explicitly in this term (see Ref.\onlinecite{verges1}).
The distortions of the six oxygens surrounding a Mn at
site ${\bf i}$ are given by the classical numbers $u_{{\bf i},\pm \alpha}$,
where in 3D (2D) $\alpha$ runs over three (two) directions $x,y$, and $z$ ($x,y$).
One widely-used simplification to the above Hamiltonian,
without losing essential physics, is to take the limit $J_{\rm H}$=$\infty$,
since in manganese oxides the Hund interaction is experimentally known to be large. In this
limit, the $e_{\rm g}$-electron spin perfectly aligns along the
$t_{\rm 2g}$-spin direction. This allows us
to reduce the first two terms of the double exchange 
Hamiltonian to:\cite{review}
\begin{equation}
H_{\rm DE}=-\sum_{\langle {\bf i,j} \rangle}[\mathcal{T}(\mathbf{S}_{\bf i},\mathbf{S}_{\bf j})
d_{\bf i}^\dag d_{\bf j} + \mbox{h.c.}],
\end{equation}
where
\begin{equation}
\mathcal{T}(\mathbf{S}_{\bf i},\mathbf{S}_{\bf j})=-t[\cos{\frac{\theta_{\bf i}}{2}}\cos{\frac{\theta_{\bf j}}{2}}
+\sin{\frac{\theta_{\bf i}}{2}}\sin{\frac{\theta_{\bf j}}{2}}\mbox{e}^{i(\phi_{\bf i}-\phi_{\bf j})}].
\end{equation}
$\theta_{\bf i}$ and $\phi_{\bf i}$ are the spherical coordinates of the core
spin $\mathbf{S}_{\bf i}$ at site
$\bf i$. The new operators $d_{\bf i}^\dag$ create an electron at site $\bf i$ with spin
parallel to the core spin at $\bf i$. Within this large Hund-coupling
approximation, the third term that produces lattice distortions becomes:
\begin{equation}
H_{\rm e-ph}=- \lambda t \sum_{{\bf i}, \alpha}(u_{{\bf i},-\alpha}-u_{{\bf i},\alpha})d_{\bf i}^\dag d_{\bf i}.
\end{equation}
This tendency is balanced by an opposite crystal elastic energy that represents
the stiffness of the Mn-O bonds:
\begin{equation}
H_{\rm ph}=t\sum_{{\bf i},\alpha}(u_{{\bf i},\alpha})^2.
\end{equation}
In the present investigations, it is important to also
consider a disorder term of the form:
\begin{equation}
H_{\Delta}=\sum_{\bf i}\Delta_{\bf i} n_{\bf i},
\end{equation}
where $\Delta_{\bf i}$ is chosen randomly from some distribution (box, bimodal,
Gaussian) and $n_{\bf i}$=$d_{\bf i}^\dag d_{\bf i}$.
Hence, the full Hamiltonian used in the present study is given by:
\begin{equation}
H=H_{\rm DE}+H_{\rm e-ph}+H_{\rm ph}+H_{\Delta}.
\label{eq:hamiltonian}
\end{equation}
In this paper, we have considered two types of disorder: (i) a binary or bimodal
distribution where
$\Delta_{\bf i}$=$\pm \Delta$, with $\Delta$ being a constant, and (ii) a box
distribution where
$\Delta$ can take any value in some interval $[-\Delta^\prime, \Delta^\prime]$.
In order to compare the results between these two distributions, one should establish
a relation among them. This can be done by requiring that the standard deviation
$\Delta x$ should be the same for both. Since $\langle
x \rangle$=$0$, one has:
\begin{equation}
(\Delta x)^2= \langle x^2 \rangle =\int_{-\infty}^{\infty}dx\mbox{ }x^2 P(x),
\end{equation}
where $P(x)$=$\delta(x+\Delta)+\delta(x-\Delta)$ for the binary distribution and
$P(x)$=$1$ in the interval $x\in [-\Delta^\prime, \Delta^\prime]$ for the
box distribution. This gives
$\langle x^2 \rangle $=$ 2\Delta^2$ for the former,
and $\langle x^2 \rangle$=$2\Delta^{\prime 3}/3$ for the latter.
Hence, the relation between $\Delta$ and
$\Delta^\prime$ should be
$\Delta^{\prime}$=$(3\Delta^2)^{\frac{1}{3}}$.

The model used in this paper contains only one orbital in the
$e_{\rm g}$ sector, instead of the more realistic case of two active orbitals.
Since using two orbitals increases substantially the CPU time, in this
first analysis the focus is on the one-orbital situation. Also note that
Motome and collaborators have studied a similar model,\cite{motome} 
as remarked in the
introduction, but they implemented the cooperative effect between oxygens
(namely the effect by which a given oxygen is shared by two octahedra)
in a different manner. This does not represent a problem since our
results for the phase diagram and density-of-states 
are very similar to those of Ref.\onlinecite{motome}.  
Finally, the technique to be used involves
the exact diagonalization of the fermionic sector, while for the classical
degrees of freedom, a standard Metropolis is used. Additional details of
the numerical methods are not presented here since they have been extensively discussed
in previous literature.\cite{review,book}

\section{Insulator-to-Metal Transition Induced by Quenched Disorder}

\subsection{Phase Diagram}

The phase diagram of the model discussed in the previous section
was investigated using spin-spin and density-density real-space
correlation functions, as well as spin and charge structure functions in
$\bf q$-space, which are
defined, respectively, as follows:

\begin{equation}
S(\mathbf{x})=\frac{1}{N}\sum_{\mathbf{i}} \langle \mathbf{S_i\cdot S_{i+x}}\rangle,
\label{eq:sx}
\end{equation}

\begin{equation}
n(\mathbf{x})=\sum_{\mathbf{i}} \langle n_{\mathbf{i}}\rangle \langle n_{\mathbf{i+x}} \rangle
- \frac{1}{N}(\sum_{\mathbf{i}}\langle n_{\mathbf{i}}\rangle)^2,
\label{eq:nx}
\end{equation}

\begin{equation}
S(\mathbf{q})=\sum_{\mathbf{i,j}}\langle \mathbf{S_i\cdot S_j} \rangle
\mbox{e}^{i\mathbf{q\cdot (r_i-r_j)}},
\label{eq:sq}
\end{equation}

\begin{equation}
n(\mathbf{q})=\sum_{\mathbf{i,j}}\langle n_{\mathbf{i}}\rangle
\langle n_{\mathbf{j}} \rangle
\mbox{e}^{i\mathbf{q\cdot (r_i-r_j)}}.
\label{eq:nq}
\end{equation}
In all of the correlations, $\langle \mathcal{O} \rangle$ stands
for the thermal average of the
operator $\mathcal{O}$ and is defined by:
\begin{equation}
\langle \mathcal{O} \rangle =
\frac{\mbox{Tr}[\mathcal{O}e^{-\beta H}]}{\mbox{Tr}[e^{-\beta H}]},
\end{equation}
where $\beta$ is the inverse temperature, $N$ is the total number of
sites, and the rest of the notation
is standard.\cite{comment-density}
In the following sections, results are given
in units of $t$, unless otherwise stated. In addition, all the
results shown in this manuscript correspond to the case of half-filling $n$=$0.5$, where
in average there is one electron for every two sites. This was the
electronic density studied in Refs.~\onlinecite{aliaga,motome}, and typically
such density is easily analyzed numerically (contrary to
other cases where more complicated charge patterns occur,
requiring larger lattices). Moreover, many real manganites have
$n$=$0.5$.

As a criteria to decide whether a particular ordered
phase is stable, namely whether long-range order is present,
the equal-time spin and charge correlation functions were used.
More specifically, a charge-ordered (CO) phase is here defined as stable if:
(i) the real-space charge correlation $n(\mathbf{x})$ at the maximum
distance possible in the finite-size clusters studied is at least
$5\%$ of the maximum value found at distance zero $n({0})$. The $5\%$
criteria is certainly arbitrary, nevertheless it is qualitatively
representative of
the physics under investigation here. Other criteria, i.e. $10\%$ or
$1\%$, simply lead to phase diagrams slightly shifted up or down
in temperature.
In addition, (ii) $\max n(\mathbf{q})$ must have its
maximum value at $\mathbf{q}$=$(\pi,\pi)$, for a checkerboard charge-ordered
state to be defined as stable. These two
criteria give very similar results in practice. For
the ferromagnetic (FM) phase, a similar definition was used:
in this case a FM phase is said to exist when $S(\mathbf{x})$ at the maximum
distance is larger than $1\%$ of $S({0})$, which is already normalized to one.
In addition, the spin correlation in momentum space must have a robust peak at (0,0).
The excellent agreement found here with previous 
investigations \cite{aliaga,motome} leads us to 
believe that our criteria
for the determination of the phase diagram is robust, and variations around
this criteria will not change qualitatively the main conclusions.

In the actual Monte Carlo simulations, the spin and pho\-no\-nic
configurations, both assumed to be classical degrees of freedom,
are updated according to the Metropolis algorithm. For the
8$\times$8 lattice, the number of Monte Carlo sweeps per site is
taken as 10$^4$ for thermalization, typically followed by another
10$^4$ sweeps or more for measurements. The simulations were done
in a two-fold way. In the first case, the starting configuration
of classical spins and oxygen coordinates was chosen to be random,
namely a high temperature configuration is selected, followed by a
slow cooling down. In the second case, simulations started with an
ordered stated at low temperatures, which is followed by a slow
heating up. It was observed that both procedures gave fairly
consistent results, and the fluctuations in the values of critical
temperatures are contained in the error bars shown.

\begin{figure}
\centering{
\includegraphics[clip,width=7.0cm]{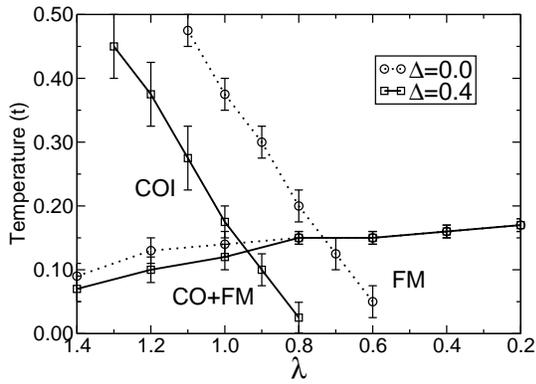}}
\caption{Monte Carlo phase diagram of the $J_{\rm H}$=$\infty$
one-orbital model as given by the Hamiltonian in
Eq.~\ref{eq:hamiltonian}, for an 8$\times$8 lattice, and overall
density $n$=$0.5$. Results at the values of disorder indicated are shown. 
The error bars are representative of the different critical temperatures
obtained using different starting configurations, as discussed in
the text. The number of quenched disorder configurations
used for the averages
varies from point to point, but the smoothness of the results
indicates that there is no crucial dependence with that number.
}
\label{fig:phase8x8}
\end{figure}

Using the criteria outlined in the previous paragraphs, the phase
diagram of model Eq.~\ref{eq:hamiltonian} is shown in
Fig.~\ref{fig:phase8x8}. In the clean limit, two phases compete: a
charge-ordered phase with a checkerboard arrangement of charge (to
be expected in the large $\lambda$ limit), and a ferromagnetic
phase at small $\lambda$ which is caused by the double exchange
mechanism.\cite{book} The existence of these phases and shape of
the phase diagram is in excellent agreement with the results of
Ref.~\onlinecite{motome} (see also Refs.~\onlinecite{aliaga,verges1}) 
although different criteria to define
long-range order were used, independent programs were written, and
even the actual models differ in the form in which cooperation is
introduced. In the clean limit there is a region where
both orders coexist at low temperature, leading to tetracritical
behavior. Other investigations of alternative models for manganites
produced first-order transitions between
competing phases with AF and FM
characteristics.\cite{review,aliaga,burgy2} In the present analysis, an AF state is not
present and phase competition is probably not so intense as in
more realistic cases.

The most interesting results arise upon the introduction of disorder in the form
of random local energies.
In this case, it was observed that the FM correlations are not much affected
by disorder, but the charge correlations are drastically modified. At
$\lambda$ in the range $0.6$ to $0.8$ and low temperatures, a CO state in the
clean  limit becomes a charge-disordered state for nonzero $\Delta$ (strength
of the on-site disorder). This is suggestive of an insulator to metal
transition. The phase
diagram and prediction of insulator to metal transition based
on the study of the density-of-states was previously discussed in
Ref.~\onlinecite{motome} (see also Ref.~\onlinecite{aliaga})
and our investigations confirm
their results. In addition, studies of conductances
discussed in the rest of the paper confirm the prediction of an
insulator to metal transition triggered by disorder. It is interesting
to remark that the FM state does not seem to play any important role
in this study, it is the insulating CO-state that is highly susceptible
to the introduction of disorder. Thus, the actual competition at the
heart of the phenomenon discussed here appears to be among charge-ordered and
charge-disordered states, as explained more extensively in the 
discussion below using a model
where the spin degree-of-freedom is removed.

\begin{figure}
\centering{
\includegraphics[clip,width=8.6cm]{\mypath{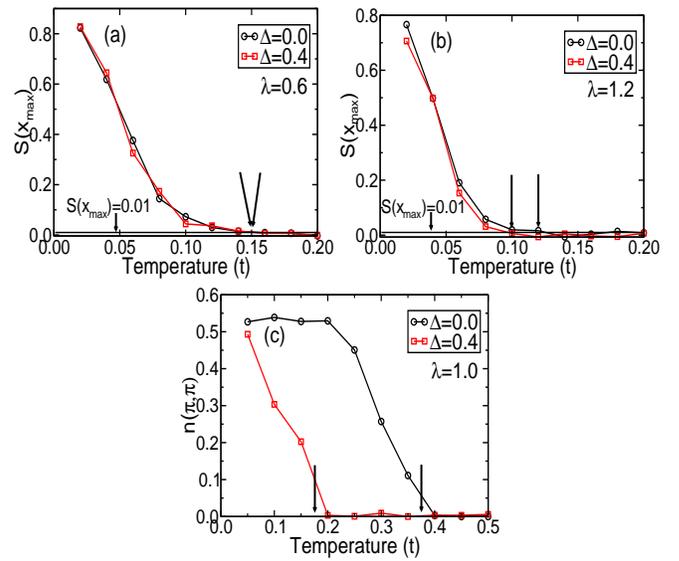}}}
\caption{The criteria used for the determination of the Curie temperature
 $T_{\rm c}$: spin-spin
correlation function at
maximum distance for the cluster studied vs. 
temperature, for $\lambda$=$0.6$ (a) and $\lambda$=$1.2$ (b).
The $S(x_{max})$=$0.01$ line is $1\%$ of its maximum value, the criteria used here
to determine FM critical temperatures (indicated by the arrows).
For $T_{\rm CO}$, $n(\pi,\pi)$ vs. temperature for $\lambda$=$1.0$ is shown in (c),
for two strengths of disorder $\Delta$, and the approximate critical temperatures
of the CO-state are indicated.
The figures show results for one disorder
configuration.}
\label{fig:sxnx}
\end{figure}

Figure~\ref{fig:sxnx} illustrates some of the evidence gathered for
the construction of the phase diagram Fig.~\ref{fig:phase8x8}.
In parts (a) and (b) the spin correlations are shown, at two
different values of $\lambda$, as indicated. In both cases, the
behavior of the spin correlations is quite similar before and
after disorder is introduced. It appears that this  degree of freedom is
not the crucial for the understanding of the phenomena, 
since it barely changes with disorder.
More relevant is the study of charge ordering, since its
associated critical temperature $T_{\rm CO}$ is substantially
modified at nonzero $\Delta$. One typical example is shown in
Fig.~\ref{fig:sxnx}(c): here the charge structure factor at
wavevector $(\pi,\pi)$ is shown. There is a clear difference
between the cases with and without disorder, with a reduction of
the critical temperature by roughly a factor two. With hindsight
based on the discussion that is presented in the next section, it is
understandable that on-site random energies will affect severely a
CO state. If these random energies are negative at some of the sites
that were not originally populated in the CO state, likely a
transfer of charge to those sites will be induced. The reciprocal
occurs with positive random energies in sites that were populated
in the clean limit. Thus, the phenomenon presented here is
expected to be basically related to charge ordering (not spin) and
also to depend on
the particular form of quenched disorder used, as confirmed later
in the discussion.

\begin{figure}
\centering{
\includegraphics[clip,width=8cm]{\mypath{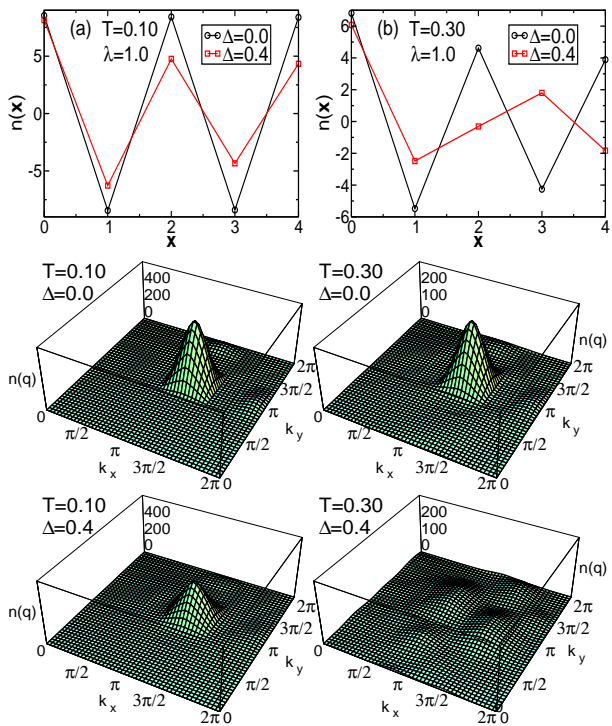}}}
\caption{Breakdown of the charge-ordered state with disorder:
For small temperatures, $T$=$0.10$, the CO phase survives
the disorder [(a),(c) and (e)]. However, at larger temperatures, $T$=$0.30$,
the initial CO state melts with disorder [(b),(d),
and (f)].}
\label{fig:nxnq}
\end{figure}

For completeness, results for just one quenched-disorder
configuration of random energies are shown in Fig.~\ref{fig:nxnq}.
Part (a) corresponds to a temperature where both with and without
the disorder (strength indicated), a CO state is
stabilized. The charge staggered pattern is robust, even
with disorder incorporated. This is caused by the cooperative
nature of the lattice distortions.
On the other hand, at $T$=$0.30$ (part (b)) the charge staggered
pattern is substantially distorted with disorder. This fact is
also very clearly shown in the charge structure factor (part
(c)) where a robust $(\pi,\pi)$ peak is replaced by a fairly
featureless background at $T$=$0.30$ and $\Delta$=$0.4$. For 
each individual configuration of disorder, the charge appears
to order following a complicated pattern (b). However, the average
over disorder produces a featureless state.

\subsection{Density-of-States}

In addition to the equal-time correlations used in the previous
section to establish the phase diagram, in this effort the
density-of-states (DOS) was also calculated. The procedure to
calculate dynamical quantities involves the full diagonalization
of the electronic problem for each configuration of classical
spins and phonons. Details can be found in Ref.~\onlinecite{book}.

\begin{figure}[h]
\centering{
\includegraphics[clip,width=7.0cm]{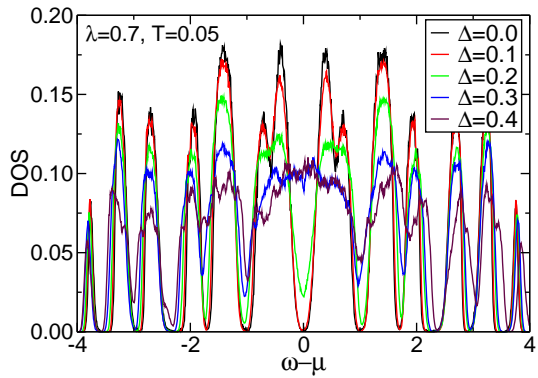}}
\caption{Density-of-states obtained with Monte Carlo
simulations at the values of $\lambda$ and temperature indicated.
The lattice is 8$\times$8 in size, and  averages over five disorder
configurations are presented (the disorder was generated with a bimodal
distribution in this example). $10^4$ thermalization steps were used,
with $10^4$ measurements. Values of $\Delta$ are indicated.
Similar results were also obtained using 4$\times$4 clusters (not shown).
}
\label{fig:bidos}
\end{figure}

Interesting results are shown in Fig.~\ref{fig:bidos},
illustrating one of the main conclusions of the present
investigation, and confirming the predictions of Ref.~\onlinecite{motome}. 
As expected, the clean-limit case $\Delta$=$0$
presents a DOS which has a \emph{gap} at the chemical potential,
in agreement with the results of Fig.~\ref{fig:phase8x8} that
locate $\lambda$=$0.7$ and low temperature within the CO phase.
The presence of oscillations at nonzero values of $\omega$-$\mu$
is due to well-known finite-size effects: a relatively small
cluster with periodic boundary conditions typically has a shell
structure in their ladder of states. These spurious oscillations do not affect the
conclusions of our study. The main result related with
Fig.~\ref{fig:bidos} is the interesting dependence of the DOS with
the magnitude of $\Delta$. In fact, at $\Delta$=$0.3$ and higher
the original CO gap at the chemical potential has simply vanished,
suggesting the instability of the insulating CO-state toward a novel state
(presumably metallic) induced by quenched disorder. 
Understanding the origin of this unexpected 
insulator-to-metal transition, triggered by disorder, will occupy most of the
rest of the present paper.

\begin{figure}[ht]
\centering{
\includegraphics[width=7.0cm]{\mypath{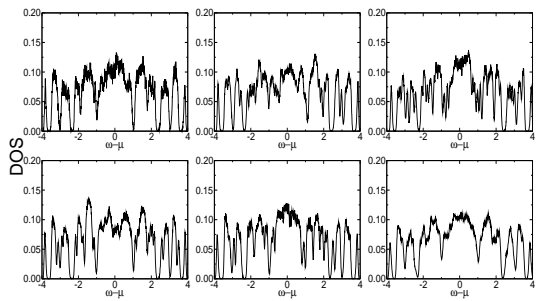}}}
\caption{First five figures correspond to the DOS obtained with
five different quenched-disorder configurations, at the same
$\lambda$, lattice size, and  temperature used in
Fig.~\ref{fig:bidos}. A bimodal distribution of disorder was
employed and results for $\Delta$=$0.4$ are shown. 
All five results are qualitatively similar. The last figure is the
average over all five.} \label{fig:average}
\end{figure}

To convince the reader that the apparently small number of
disorder realization used in Fig.~\ref{fig:bidos} is sufficient
for our purposes, in Fig.~\ref{fig:average} results for five
independent quenched-disorder configurations are shown.
Qualitatively, the five present states at the chemical potential
as observed in the average Fig.~\ref{fig:bidos}.

\begin{figure}[ht]
\centering{
\includegraphics[clip,width=7.0cm]{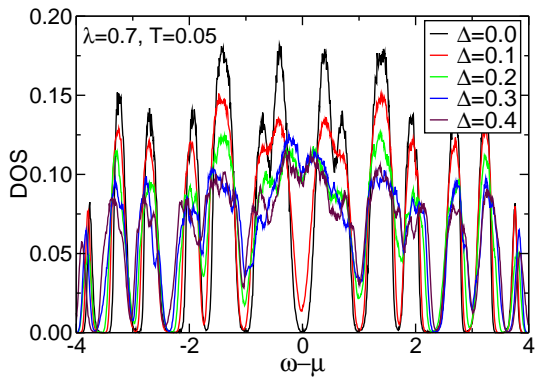}}
\caption{Density-of-states (averaged over five disorder configurations) using
a box distribution (instead of bimodal).
10$^4$ thermalization steps were used, with  10$^4$ sweeps for
measurements.}
\label{fig:boxdos}
\end{figure}

To analyze the dependence of the results with the type of disorder
introduced, in Fig.~\ref{fig:boxdos} results for a box
distribution, instead of bimodal, are shown at the couplings
indicated. Once again, the DOS presents a CO state at small
$\Delta$, which is replaced by a metallic-like DOS when $\Delta$
is increased.

\begin{figure}[ht]
\centering{
\includegraphics[clip,width=7.0cm]{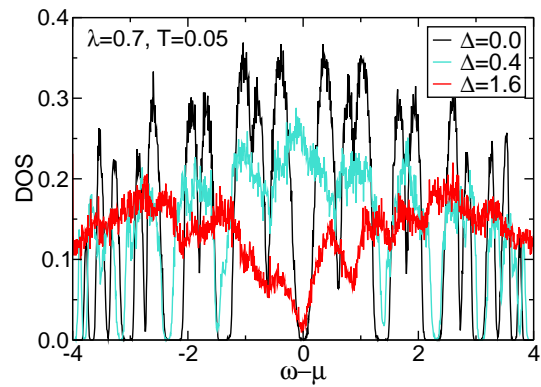}}
\caption{The "closing" and "re-opening" of the gap in the density-of-states
as the disorder strength $\Delta$ is increased, using a bimodal distribution.
Temperature, $\lambda$, and $\Delta$'s are indicated.}
\label{fig:dosbi}
\end{figure}

To study the dependence of the results with $\Delta$ over a wider
range, in Fig.~\ref{fig:dosbi} results for $\Delta$ as large as $1.6$
are shown, using a bimodal distribution. At large $\Delta$, it is
natural to expect that the states at the Fermi level will not be
filled, since this strong disorder {\it must} induce localization of
charge at the sites where the on-site random energy happens to be
large and negative. Then, the DOS presents an interesting
non-monotonic behavior with an insulator at both $\Delta$ small and large,
while the system is metallic at intermediate $\Delta$.

\begin{figure}
\centering{
\includegraphics[clip,width=7.0cm]{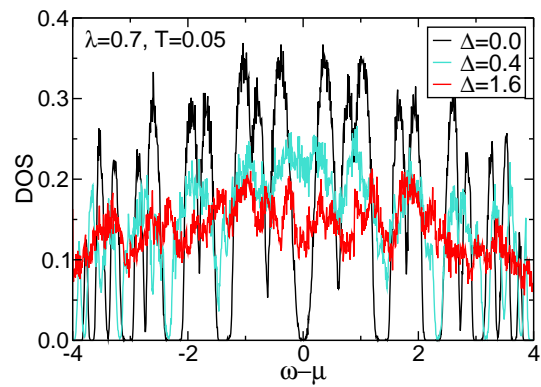}}
\caption{Analogue of Fig.~\ref{fig:dosbi} but for a box
distribution. In this case, the gap that closes at intermediate
$\Delta$ does not form again for larger values of the disorder
strength.} \label{fig:dosbox}
\end{figure}

The situation is different for the box distribution. In this case,
it is expected that large $\Delta$ will not alter the results so
dramatically as for the case of the bimodal distribution, since
there are always random energies that will happen to be small and
they do not lead to localization. This is confirmed by the results
of the simulation in Fig.~\ref{fig:dosbox}, showing that
states are present at the Fermi level even for large disorder.

\subsection{Landauer Conductance}

\begin{figure}[h]
\centering{
\includegraphics[clip,width=5.5cm]{\mypath{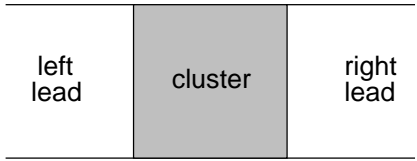}}}
\caption{Geometry used for the calculation of the conductance.
The interacting region (cluster) is connected by
ideal contacts to semi-infinite ideal leads. The configurations
of classical localized spins and phonons are produced by Monte Carlo simulations.}
\label{fig:leads}
\end{figure}

While the presence of weight at the Fermi level in the DOS is suggestive of a
metallic state, it may occur that those states are not extended but
localized. To clarify this issue
the conductance, $G$, of clusters described by the Hamiltonians used
in this effort was also evaluated. The calculation was carried out
using the Kubo formula adapted to geometries  usually employed in the
context of mesoscopic systems.\cite{verges} The actual expression for $G$ is
\begin{equation}
G=2\frac{e^2}{h} \mathrm{Tr}\left[(i\hbar \hat{v}_x) \mathrm{Im} \hat{\mathcal{G}}(E_{\rm F})
 (i\hbar \hat{v}_x) \mathrm{Im} \hat{\mathcal{G}}(E_{\rm F})\right],
\label{eq:conductance}
\end{equation}
where $\hat{v}_x$ is the velocity operator in the $x$ direction (assuming current
flows along that direction) and
$ \mathrm{Im} \hat{\mathcal{G}}(E_{\rm F})$ is obtained from the advanced and retarded
 Green functions using
$2i \mathrm{Im} \hat{\mathcal{G}}(E_{\rm F})$=$\hat{\mathcal{G}}^R(E_{\rm F})-
\hat{\mathcal{G}}^A(E_{\rm F})$, where $E_{\rm F}$
is the Fermi energy. The cluster is considered
to be connected by ideal contacts to
two semi-infinite ideal leads, as represented in Fig.~\ref{fig:leads}.
The current is induced by an infinitesimal voltage drop.
 This formalism avoids some of the 
problems associated with finite systems with periodic boundary conditions, such as
 the fact that the optical conductivity is given by a
sum of Dirac $\delta$ functions due to the discrete nature of
the matrix eigenvalues. 
Finding a zero-frequency Drude peak with finite weight
corresponds, in principle, to an ideal
metal -- i.e. a system with $zero$ resistance -- unless 
an arbitrary width is given
to the zero frequency $\delta$-function. In addition, in numerical studies sometimes
it occurs that the weights
of the zero-frequency delta-peak are {\it negative} due to subtle finite-size
effects.\cite{RMP} For these reasons, calculations of DC resistivity using finite
closed systems are rare in the literature. All these
problems are avoided with the formulation described here, which can be readily
applied to the Hamiltonians employed in our study where the fermionic sector
is quadratic. The problem basically amounts to the calculation of
transmissions across nontrivial backgrounds of classical spins and lattice distortions.

In practice, the entire equilibrated cluster, as obtained
from the MC simulation, is introduced in the
geometry of Fig.~\ref{fig:leads}. The ideal leads enter the formalism through
self-energies at the left/right boundaries, as described in Ref.~\onlinecite{verges}.
In some cases a variant of the method, explained also
in Ref.~\onlinecite{verges}, was
used to calculate the conductance:
Instead of connecting the
cluster to an ideal lead with equal hoppings, a replica of the cluster
was used at the sides. This method, although slightly more CPU time consuming,
takes into account all of the Monte Carlo data for the cluster,
including the periodic boundary conditions.
The final step in the calculation is to carry out averages over the localized spin
configurations and oxygen coordinates provided by the Monte Carlo procedure. The
physical units of the conductance $G$
in the numerical simulations for 2D are $e^2/h$ as can be
inferred from Eq.~\ref{eq:conductance}. In 2D the conductivity $\sigma$
coincides with the
conductance, while in 3D $\sigma$ is simply given by the
conductance divided by the linear size of the lattice, assumed cubic.

\begin{figure}[h]
\centering{
\includegraphics[clip,width=7.0cm]{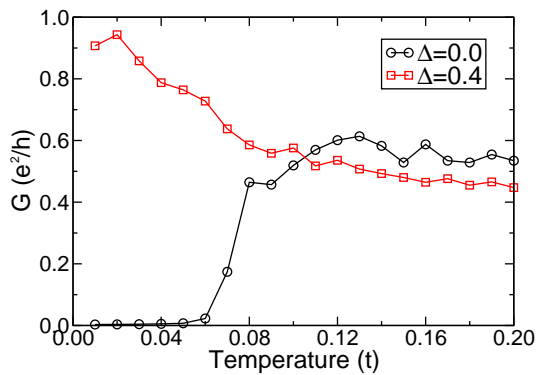}}
\caption{Conductance (e$^2/h$) vs. temperature for the one-orbital
model with cooperative phonons at $\lambda$=$0.7$ and using a
12$\times$12 cluster, in the geometry shown in
Fig.~\ref{fig:leads}.} \label{fig:convst}
\end{figure}

\begin{figure}
\centering{
\includegraphics[clip,width=7.0cm]{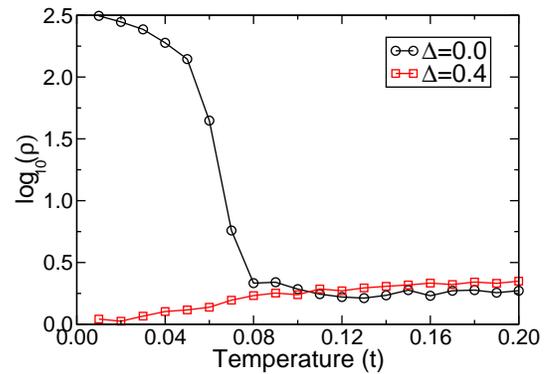}}
\caption{Logarithm (in base 10) of resistivity (inverse of conductivity) vs.
temperature for the same lattice and couplings used in
Fig.~\ref{fig:convst}. Note the similarity of the results with
those obtained varying magnetic fields near AF-FM first-order transitions 
(see Ref.~\onlinecite{aliaga}).} \label{fig:resvst}
\end{figure}

A representative result of our effort is shown in
Fig.~\ref{fig:convst}, where $G$ vs. temperature is shown at an
electron-phonon coupling and disorder strength where the DOS study
of the previous section revealed the insulator to metal
transition. In the clean-limit, the nearly vanishing value of $G$
at low temperatures reveals the presence of an insulator, which it
is known to be a charge-ordered state. The transition to a
disordered state occurs with increasing temperature at
approximately $T$=$0.08$-$0.10$ ($\Delta$=$0$), value compatible but
slightly lower that found through the charge correlations (this
may simply be due to the different lattice sizes used). The most
remarkable result is shown at $\Delta$=$0.4$. Here, the
conductance is very similar to the clean-limit result at
temperatures above the transition, but at low temperatures they
are drastically different. Compatible with the finite DOS at the
Fermi level found in the previous section, the system with nonzero
quenched disorder has a {\it finite conductance} at low-$T$, which
actually increases with decreasing temperature as in a metallic
state. The inverse of the conductance (i.e. the resistivity in 2D)
vs. $T$ is shown in Fig.~\ref{fig:resvst}, using a logarithmic
scale to amplify the nearly vanishing conductance of the
insulating state of Fig.~\ref{fig:convst}. The
introduction of quenched disorder transforms an insulator in the
clean limit to a metallic state in the dirty case. As a
consequence, the states at the Fermi level induced by disorder
that were observed in the previous section indeed correspond to
extended states that transport charge, a very interesting result.
Note, however, that the conductance in the metallic state is
substantially smaller than its maximum value (10 for a
12$\times$12 cluster), indicative of a {\it dirty metal}. This result is
also compatible with experiments in manganites that have revealed
similar characteristics.\cite{tokura,book} Then, the metallic state is far from
perfect and there must exist sources of scattering that prevents the
smooth flow of charge. In subsequent sections, it will be argued that this metallic
state is percolative-like, result compatible with its relatively
high resistance.

\begin{figure}[ht]
\centering{
\includegraphics[clip,width=7.0cm]{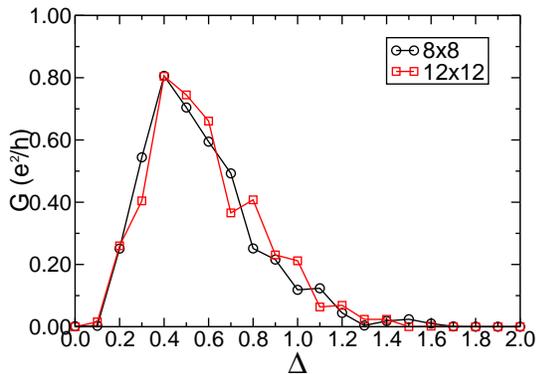}}
\caption{Conductivity vs. disorder strength for model
Eq.~\ref{eq:hamiltonian}, at $\lambda$=$0.7$ and $T$=$0.05$. $10,000$
thermalization and $2000$ measurement Monte Carlo sweeps were used.
Results for two lattice sizes are shown, suggesting 
that finite-size effects could be small. } \label{fig:realconvsd}
\end{figure}

In Fig.~\ref{fig:realconvsd}, results for the conductance at low
temperature as a function of $\Delta$ are shown for two different
cluster sizes. The size effects appear to be very small, although
the clusters studied are not sufficiently large to show
convincingly that a metallic state was indeed stabilized. The
behavior as a function of $\Delta$ reveals a range of insulating
behavior at small $\Delta$, followed by a well-defined peak in the
conductance at intermediate values -- where $\Delta$ is comparable
to the gap, as explained later in the text -- followed by a subsequent
decrease of $G$ at $\Delta$ of order 1 or larger. This 
large-$\Delta$ behavior can easily be understood due to strong
localization of carriers in that limit. However, the presence of a
metallic state at intermediate $\Delta$ -- as opposed to a smooth
interpolation between the two insulating limits at small and large
$\Delta$ -- is puzzling. Its understanding is the main goal of the
next section.

\section{Results for a simplified model}

The results presented in the previous section have revealed an
insulator to metal transition induced by on-site random-energy 
disorder, when the clean
limit starting state is charge ordered. This conclusion
appears clearly both in the DOS and in the conductance, and it is
already evident in lattices as small as 4$\times$4. Then, the mechanism
must be easy to understand since it cannot depend on subtleties related
with the bulk limit or large correlation lengths. In order to
clarify the origin of the insulator-metal transition, here a ``toy
model'' involving only electrons will be introduced, and shown to
behave quite similarly as the actual ``real model'' that has
electrons, localized spins, as well as lattice degrees of freedom. The
definition of the model is:
\begin{equation}
H=-\sum_{\langle {\bf i,j} \rangle}t_{\bf ij}(d_{\bf i}^\dag d_{\bf j} +
\mbox{h.c.}) +\sum_{\bf i}\alpha_{\bf i}n_{\bf i}+\sum_{\bf i}\Delta_{\bf i}n_{\bf i}.
\label{eq:toyh}
\end{equation}

This simplified toy model includes a hopping term for spinless
fermions (first term), as in the realistic case at 
large Hund coupling. The hopping amplitude
$t_{\bf ij}$ will first be considered constant and equal to $t$ in
this section, but later a random hopping will also be studied
for completeness. The most important simplification of
Eq.~\ref{eq:toyh}, as compared to Eq.~\ref{eq:hamiltonian}, is the
following: instead of lattice distortions to induce charge
ordering, a mere modulation of the chemical potential is
introduced ``by hand'' (second term in Eq.~\ref{eq:toyh}), where
$\alpha_{\bf i}$=+$\alpha$ and -$\alpha$ are located on a checkerboard
arrangement. This is sufficient to generate the analog of the CO state induced
by phonons, namely a charge modulation leading to a
checkerboard pattern. In all the toy model 
simulations reported
in this paper $\alpha$ was kept equal to $0.5$: for computational
simplicity it was
not modified to mimic different values of $\lambda$ and concomitant
charge gaps. However, the qualitative results emerging from the analysis
 are sufficiently
clear, that there was no need to further consider
$\alpha$ as a tuning variable. Finally, to incorporate the
quenched disorder, an on-site random-energy term as in the realistic case 
was included as well.

\begin{figure}[h]
\centering{
\includegraphics[clip,width=7.0cm]{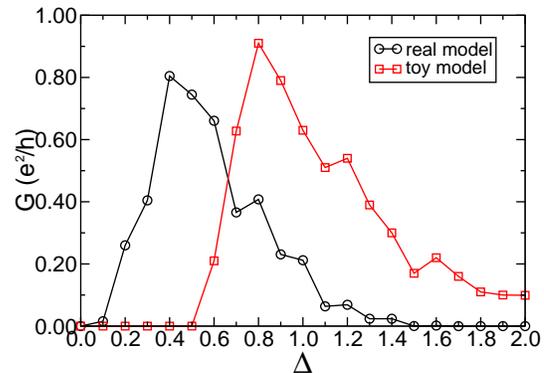}}
\caption{Conductivity vs. disorder strength $\Delta$, comparing
results for the ``real'' phononic model Eq.~\ref{eq:hamiltonian} at
$\lambda$=$0.7$ against the ``toy'' model. In both cases a bimodal
distribution is used, and results are shown using a 12$\times$12
lattice at $T$=$0.05$. In the calculation of the 
conductance $G$ for the real
model, $10,000$ sweeps for thermalization and $2000$ for measurements
were used. For the toy model the curve shows an average over $100$
configurations of disorder, and $\alpha$ is $0.5$. The different locations
of the conductance peaks can be brought to the same value
by modifying $\alpha$, an unnecessary task in our qualitative study.}
\label{fig:conrealvstoy}
\end{figure}

\begin{figure}[ht]
\centering{
\includegraphics[clip,width=7.0cm]{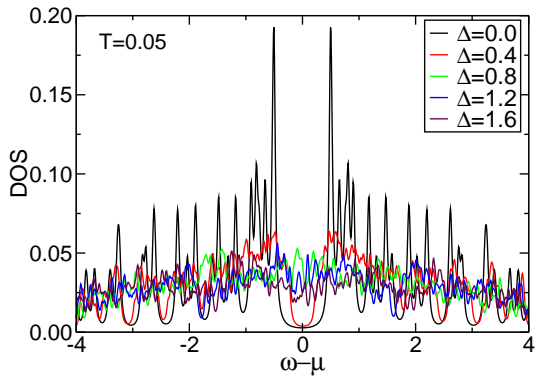}}
\caption{Density-of-states of the toy model Eq.~\ref{eq:toyh} for
several values of the disorder strength $\Delta$. The lattice is
$16\times16$ and temperature $T$=$0.05$. At the Fermi
level states are generated, as it occurs in the realistic model
studied in previous sections.} \label{fig:toydos}
\end{figure}

The calculation of conductances vs. disorder strength, as well as
other observables, proceeds straightforwardly for the toy model as
a special case of the realistic one. There is no need to carry out
any Monte Carlo simulation for the toy model, the results are exact.
A remarkable result is shown
in Fig.~\ref{fig:conrealvstoy}, where $G$ vs. $\Delta$ is shown
for both the realistic and the toy models, on the same lattice and
at the same temperature. Besides a harmless shift in the location
of the peak feature, the shape of the curve is basically the same,
revealing insulator-metal-insulator characteristics with
increasing $\Delta$ in both cases. Then, the simplified toy model
is sufficiently robust to have the same properties as the real
model, and its analysis will likely unveil the origin of the
effect. This is confirmed by a study of the DOS as well
(Fig.~\ref{fig:toydos}), since the gap in the clean-limit caused
by the staggered-modulation in the local-energy induced by the
second term of Eq.~\ref{eq:toyh} is filled by the introduction of
disorder.

\begin{figure}[ht]
\centering{
\includegraphics[clip,width=7.0cm]{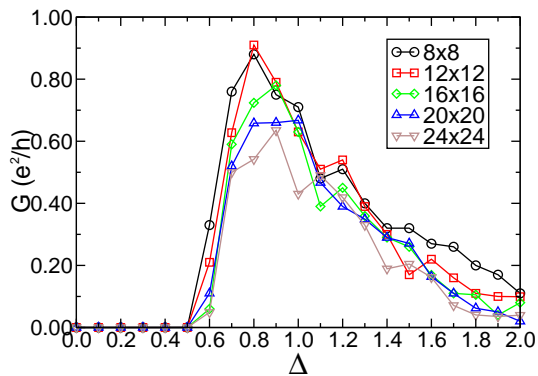}}
\caption{Conductivity vs. disorder strength for the toy model, averaged over
$100$ disorder configurations at $T$=$0.05$. Results for several cluster sizes
are shown. In 2D, conductivity and conductance are the same.
The size effects appear to be mild, but the slight reduction of $G$ with
lattice sizes in the central region could still lead to an insulating state
in the bulk limit. More work is needed to fully understand this subtle issue.}
\label{fig:norcon}
\end{figure}

The analysis of the real model -- i.e. including phonons and
localized spins as active degrees of freedom -- was limited to
12$\times$12 lattices and a few realizations of disorder. With the
toy model, a far better study can be conducted. For example, on
Fig.~\ref{fig:norcon} an analysis of $G$ vs. $\Delta$ at several
lattices is shown. All of the curves have the same shape, and the
size effects are moderate. However, there is a slow
decrease in the conductance at the peak. Given the error bars of
the simulation (the oscillations in the results between subsequent
$\Delta$'s can provide a crude estimation of these error bars),
and the rapid growth of CPU time with increasing size, an even
more careful analysis cannot be carried out. As a consequence, it
is not possible in this early investigation to be conclusive about
the metallic nature of the state in the bulk 2D limit,
particularly in view of the subtleties that are sometimes related
with similar studies in the case of Anderson weak-localization effects
(this particular issue is quantitatively discussed in the Appendix).
Nevertheless, the qualitative analysis to be discussed below will 
be sufficient to understand the nature of the metallic
state in finite-size clusters, as a first step toward applications
to real compounds.

\begin{figure}[h]
\centering{
\includegraphics[clip,width=7.0cm]{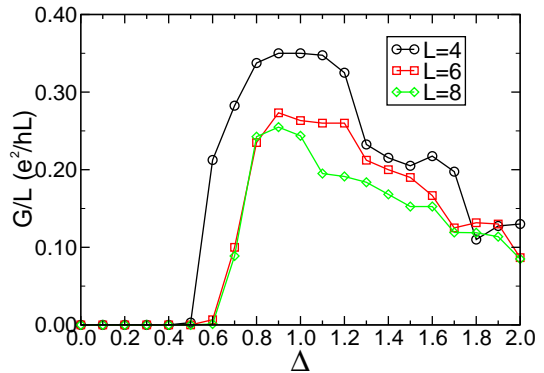}}
\caption{Conductivity (in $e^2/hL$ units)
vs. disorder strength for the toy model in
3D, averaged over 100 disorder configurations. $L$ is
the size length of a $L^3$ cube.}
\label{fig:3dcon}
\end{figure}

The situation in 3D is similar, since the results for the
conductivity ($G/L$, where the lattice size is $L^3$) of the toy
model seem to converge with increasing size of the cube used for
the simulation, as shown in Fig.~\ref{fig:3dcon}. While it is not
impossible that the conductivity could slowly decrease to zero
as $L$$\rightarrow$$\infty$, the results are suggestive that a metallic state
could  indeed be found in the bulk 3D limit. More work is needed to
fully confirm this assumption.

\section{Understanding the Metal-Insulator Transition}

\subsection{Proposed Explanation}

\begin{figure}[ht]
\centering{
\includegraphics[clip,width=7.3cm]{\mypath{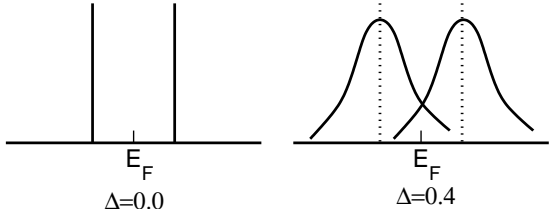}}}
\caption{Cartoonish view of the generation of weight at the Fermi level
by disorder. If in the clean limit the system is insulating (with no states available at
the Fermi level), the mere broadening of these states by disorder
certainly introduces states at $E_{\rm F}$ after some $\Delta$ strength
similar to the gap is reached. However, it remains
to be investigated whether the states at the Fermi level generated by disorder
are truly metallic, or whether they are insulating.}
\label{fig:arg}
\end{figure}

The essence to understand the insulator to metal transition found
with increasing disorder-strength appears to be very simple, and a
starting point is illustrated in Fig.~\ref{fig:arg}. Suppose that
in the clean-limit case, $\Delta$=$0$, the Fermi level is in between
two states separated by a gap (and, as a consequence, the state is
an insulator). It is natural to expect that with
 increasing on-site energy disorder those two
states will increase their respective widths -- since the on-site
random energies directly affect the energy location of the state
-- and there will be a value of $\Delta$ (of the order of the
clean-limit gap) where a robust number of states will be located
at the Fermi energy. The results we have found in our
investigation are compatible with this simple picture for the
generation of weight at the Fermi level. In this respect,
increasing temperature in the clean-limit at fixed large $\lambda$ or
increasing $\Delta$ at fixed small $T$ and large $\lambda$
appear to be analogous procedures to transform an insulator into a
metal. However, Fig.~\ref{fig:arg} can only be a rough starting
point to rationalize the metallic conductance since the presence
of states at the Fermi energy is not sufficient to guarantee that
the systems is indeed metallic. In fact, Fig.~\ref{fig:arg}
applies even in the limit of vanishing hopping, where the charge is
localized.

\begin{figure}[ht]
\centering{
\includegraphics[clip,width=8.4cm]{\mypath{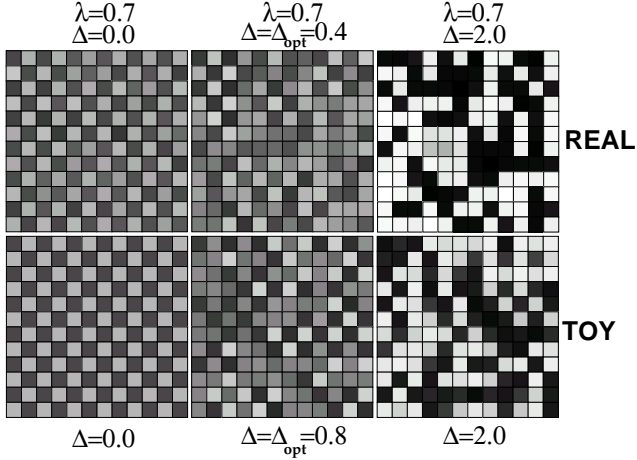}}}
\caption{Monte Carlo ``snapshots'' obtained using a 12$\times$12
cluster and temperature $T$=$0.05$ showing the electronic density
at each site in tones of black and white. Results are shown for
both the ``real'' and ``toy'' models (Eqs.~\ref{eq:hamiltonian}
and \ref{eq:toyh}, respectively). In the former, $\lambda$=$0.7$
is used. In the three cases, results for the clean limit
$\Delta$=$0$, as well as intermediate and large values of $\Delta$
are presented. Both the real and toy models at $\Delta$=$0$ have
densities approximately $0.7$ and $0.3$ in a checkerboard pattern,
while at $\Delta$=$2$ the two
clearly-distinguished densities are close to $0.95$ and $0.05$. A
detailed discussion is in the text, but the reader can already
visualize the localized nature of the charge in both limits of
small and large $\Delta$, while the snapshots at intermediate
disorder show a more disordered
distribution of populated sites, leading to a metallic state. It
is also to be remarked the similarity between the results for both
models.} \label{fig:realvstoy}
\end{figure}

The most revealing procedure to understand the I-M-I transition 
with increasing $\Delta$ found
in the real and toy models involves the study of Monte Carlo ``snapshots'' of the charge
configurations. This is useful in the real model because the classical
degrees of freedom freeze at low temperatures, and snapshots are representative
of the physics (a state with strong quantum fluctuations may have
many configurations with equal weight, complicating the analysis).
Typical snapshots are shown in Fig.~\ref{fig:realvstoy}, where
results for the electronic density of both the real and toy models
are contrasted. It is clear that in the three regimes of relevance
-- small, intermediate, and large $\Delta$ -- there is an
excellent agreement between the two cases, and the toy model
appears to capture the essence of the behavior of the more
realistic model Eq.~\ref{eq:hamiltonian}. Let us now analyze the
results. At small $\Delta$, the checkerboard pattern certainly
prevents the transmission of charge, and the system is insulating
in both cases. At large $\Delta$, the sharp color-contrast between the
sites occupied and those empty are indicative of a strong
localization of charge. Energetically this occurs to take advantage of the
regions with large and negative $\Delta$, where at every site
either occupancy zero or one is favored. This regime is insulating
as well.

\begin{figure}[h]
\centering{
\includegraphics[clip,width=7.4cm]{\mypath{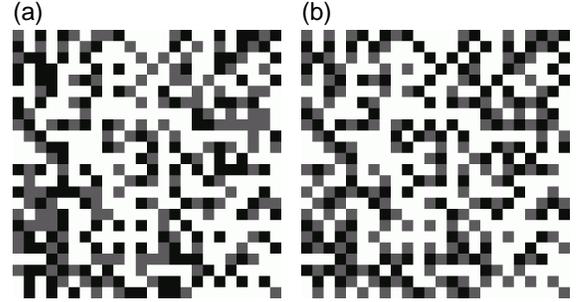}}}
\caption{Example showing that the sites with densities approximately $0.5$
are correlated with those with small effective on-site energy,
for the particular case $\Delta$=$0.8$, and $\alpha$=$0.5$, on a 24$\times$24 lattice.
On the right panel, sites with small effective on-site energy
 are shown in black if $\Delta$-$\alpha$=$0.3$ and light grey if
$\Delta$-$\alpha$=-$0.3$, the rest are in white. On
the left panel, results for the toy model are shown.
Here sites with densities between $0.50$ and $0.75$ are in black, between $0.25$
and $0.50$ in light grey, and the rest
in white. The agreement between left and right is almost perfect. This two-colors
view helps in the understanding of percolation. In two dimensions the critical percolative
fraction is $0.5$, precisely the number of sites with small effective
on-site energy. The system is then only at the $verge$
of percolation. However, in 3D it would be fully percolated since the
critical fraction is 0.25.}
\label{fig:toyvspot}
\end{figure}

The most interesting result occurs at intermediate values of
$\Delta$, in the metallic regime. Here the density is closer to
its mean value $0.5$ in several lattice sites, in spite of the
$\alpha$ modulation of the toy model. To understand the results,
consider for instance the case of a bimodal distribution for the
quenched disorder, as in Fig.~\ref{fig:realvstoy}. Considering the
second and third terms of Eq.~\ref{eq:toyh} in the toy model, then
locally the effective chemical potential can take four values,
namely $\Delta$+$\alpha$, $\Delta$-$\alpha$, -$\Delta$+$\alpha$,
or -$\Delta$-$\alpha$. The case of interest for metallicity is
obtained at $\Delta$$\sim$$\alpha$ or larger -- in excellent agreement with
the numerical results -- where  approximately half the sites have
a small effective on-site energy, and these sites are now available
for the transport of charge with average density $0.5$ (while the
other sites either have a very large or very small chemical
potential, favoring either minimum (zero) or maximum (one) local
occupancy, with both cases not useful for charge transport). The
sites with small effective on-site energy would not contribute 
to a metal if they were isolated (for instance if they were
regularly spaced), but for a {\it random} distribution of
locations a metallic state can be formed if there is a large
enough number to induce percolative-like phenomena.
Figure~\ref{fig:toyvspot} further clarifies this issue. The
calculations of conductances appear to imply that indeed such
percolation has occurred at the densities here considered. This
appears to be the essence of the insulator to metal transition.
The state that conducts appears to be highly
inhomogeneous, compatible with the low value of its conductance.
Clearly, insulating regions survive. The elastic cooperative
nature of the oxygen coordinates enlarges the regions that are
either charge ordered or disordered.

\begin{figure}[h]
\centering{
\includegraphics[clip,width=6.5cm]{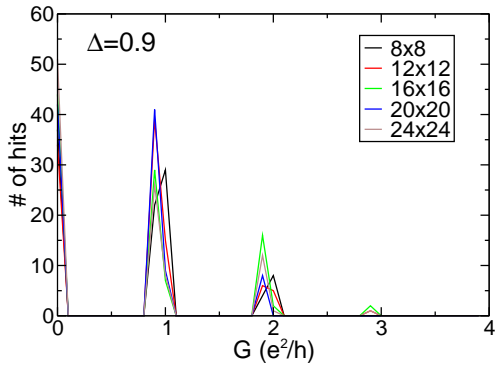}}
\caption{Histogram of conductances observed in the study
of the toy model, corresponding to $\Delta$=$0.9$, and at $T$=$0.05$,
for many values of the disorder.
The conductances
can only take integer values, here they have been broadened for a better
illustration. Size effects appear to be small. Several configurations
are insulating ($G$=$0$), while
others conducting (although poorly).}
\label{fig:histcon}
\end{figure}

To clarify further the poor nature of the metallic state, a
frequently-mentioned issue in the study of conductances is their
distribution. Sometimes the mean value, as the one discussed thus
far, is not representative due to the large width of the
distribution. However, an example of $G$-distributions is shown in
Fig.~\ref{fig:histcon}, for the toy model at a value of $\Delta$
where $G$ indicates a metallic state. The result shows that the
distribution is not abnormally wide, but the only values that
visibly appear are 0, 1, and 2, while its full possible range of
values extends up to 22 in the case of the largest lattice
studied. The distribution shown also indicates that slightly less
than half the disorder configurations used have zero
conductance, while the others have the minimum and next values,
borderline compatible with a metal. Then, the metallic
state discussed in this paper corresponds to a {\it poor metal},
and the snapshots previously discussed indicate that this property
arises from inhomogeneities. The system appears to be near percolation
and, as a consequence, there are only a limited number of paths
for conductance across-the-sample.

Although the use of standard formulas for percolative processes could be tempting (for
instance to predict the critical density of metallic sites needed to
percolate, and the corresponding critical exponents), 
here the electronic-density at each site is not simply found to
correspond to just four numbers, as the effective on-site energies
argumentation would indicate. The reason is that the kinetic
energy of electrons (first term in Eq.~\ref{eq:toyh}) must also be
considered, and the electronic density ``spills'' from favorable
to unfavorable sites if they are close to one another. This
proximity or spilling effect is not considered in traditional
models of percolation -- where links or sites are sharply either
metallic or insulating -- preventing the straightforward use of 
standard formulas deduced for classical random-resistor networks 
to the case analyzed here.

\subsection{Consequences of the Proposed Explanation}

\subsubsection{Results with Random Hoppings}

The model used both here and in the previous
in\-ves\-ti\-ga\-tion\cite{motome} employs a local random-energy term as a
source of disorder. This form of disorder appears to be crucial
for the argumentation discussed in the previous section to explain the
stabilization of a poor metallic state at intermediate $\Delta$.
The notion of an effective site-energy 
that nearly cancels at $\Delta$$\sim$$\alpha$ for
approximately half the sites is very important to render those
sites conducting. Then, the present analysis would predict a lack
of universality of the disorder-induced insulator-metal transition
upon modifications in the explicit form in which the disorder is introduced.
In particular, introducing randomness in the {\it hopping
amplitudes}, rather than in the on-site energies, should not lead to
a metal according to the previously proposed explanation of the
effect, since the effective site energies will not be affected. To
verify this idea, in the toy model the on-site disorder was
switched off, but now the hopping amplitudes were given an extra
random component. More specifically, the hoppings were modified
such that $t_{\bf ij}$=$t$+$\tau_{\bf ij}$, while $\Delta_{\rm
i}$ was switched off in Eq.~\ref{eq:toyh}. The random number
$\tau_{\rm ij}$ was chosen from a bimodal distribution with values
$\pm \tau$.

\begin{figure}
\centering{
\includegraphics[clip,width=7.0cm]{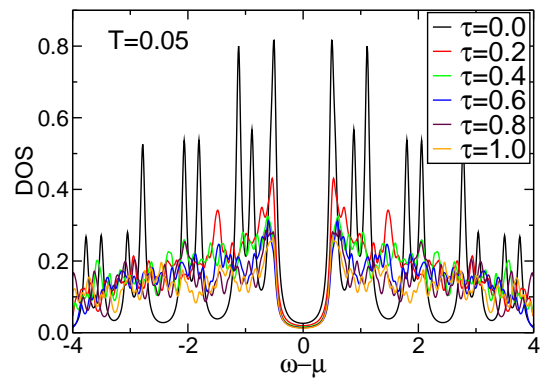}}
\caption{Density-of-states for the toy model where now the disorder is introduced
through random hoppings, rather than random on-site energies. Here, the hopping
disorder is again a bimodal function with values $\pm \tau$. The results
are at $T$=$0.05$ and obtained on a 12$\times$12 cluster. The gap does not close
under the influence of disorder in the hoppings.}
\label{fig:rantdos}
\end{figure}

The results for the density-of-states are shown in
Fig.~\ref{fig:rantdos} for several values of $\tau$. 
There is a drastic difference  when compared with the case of random
on-site energies. For random hopping amplitudes the gap
present at $\tau$=$0$ is not filled by the random hoppings with
increasing $\tau$. The state remains insulating at all values of
$\tau$, small, intermediate, and large. This is compatible with
the explanation for the insulator-metal transition discussed in
the previous subsection, which relies crucially on the effect of
on-site disorder.

\subsubsection{Simulations with Frozen Spins}

\begin{figure}
\centering{
\includegraphics[clip,width=7.0cm]{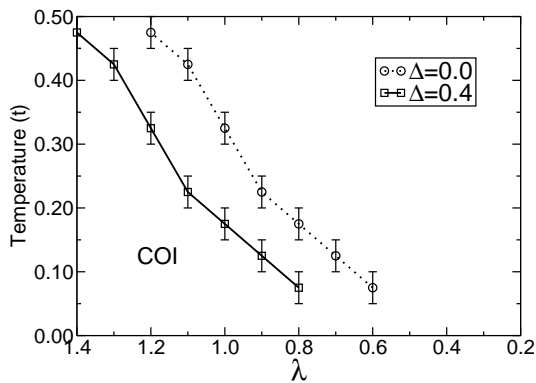}}
\caption{Phase diagram of model Eq.~\ref{eq:hamiltonian} obtained
in the limit where the localized spins are frozen in a
ferromagnetic state at all temperatures. As in the case of
non-frozen spins, $10^4$ thermalization and $10^4$ measurement
Monte Carlo sweeps were typically used at each temperature and
$\lambda$. The lattice was 8$\times$8. The criteria used to
determine the transition temperatures are the same as for the
simulations with active spins. Results both with and without
disorder are shown, with a behavior similar to that found in
Fig.~\ref{fig:phase8x8}. The phase diagram at very low
temperatures is difficult to obtain due to the rapid increase in
CPU time required to remain ergodic (configurations are separated
by large energy barriers).} \label{fig:phase}
\end{figure}

Another important prediction of the proposed explanation for the
I-M transition is the irrelevancy of the spin 
degree-of-freedom, and its concomitant FM phase. If the idea is correct,
then the presence of the CO vs. FM phase competition is not
crucial, but more important is the competition between
charge-ordered and charge-disordered states. To verify this idea a
simulation was carried out freezing the
localized spins to a perfectly ferromagnetic state at all temperatures, 
effectively
removing the spin as a degree-of-freedom. The Monte Carlo phase
diagram is shown in Fig.~\ref{fig:phase} and indeed it shows the expected
behavior for the CO state, before and after including disorder.

\begin{figure}[ht]
\centering{
\includegraphics[clip,width=7.0cm]{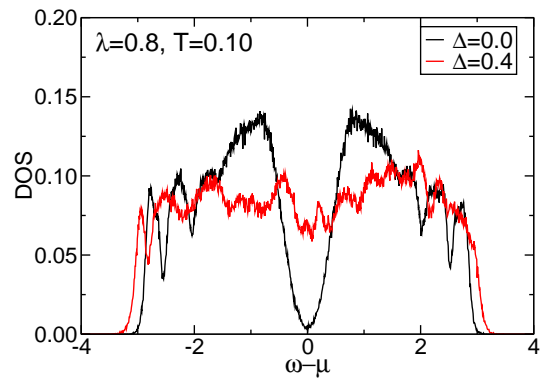}}
\caption{Density-of-states obtained using the model with frozen spins, leaving
only the charge as active degree-of-freedom. The clean-limit gap is
still found to close with disorder,
even in the absence of CO-FM competition. The lattice
is 8$\times$8.}
\label{fig:dosl08t010}
\end{figure}

The density-of-states for the model with only the charge as active
degree-of-freedom is shown in Fig.~\ref{fig:dosl08t010}, with and
without disorder. The insulator at $\Delta$=$0$ and metal
at $\Delta$=$0.4$ are present qualitatively as in
the case with active spins. It appears that the
insulator-metal transition only depends on the presence of charge
ordered and disordered states, and their competition. The
ferromagnetic component is of no consequence for
the effect discussed in this paper.

\section{CONCLUSIONS}

In this paper, an explanation was proposed
for  a recently discovered insulator-to-metal transition
induced by disorder\cite{motome} 
in a model for manganites with cooperative phonons.
The transition occurs when
the disorder is introduced in the form of random on-site energies.
Our effort presented here confirms the effect, showing that the
clean-limit gap in the density-of-states is filled by disorder. Moreover,
a calculation of the conductance was implemented, with a Landauer formulation
borrowed from the context of mesoscopic physics. This calculation confirms
that for intermediate values of the disorder strength $\Delta$ the system
is indeed metallic, at least for the clusters studied here. A better
finite-size scaling should be carried out to fully confirm the metallic
nature of the state in the bulk limit, but this huge effort is left for
future investigations. Our limited size analysis thus far indicates that the
conductivity will remain finite in the bulk limit, particularly
in three dimensions.

To understand the origin of this transition, an even simpler model
without active phononic degrees-of-freedom was proposed. This new model
behaves very similarly as the original one, with a DOS that is filled
by disorder, a quite similar conductance vs. $\Delta$ curve, and similar
Monte Carlo snapshots. Both this model and the original one, present
a percolative-like behavior in the metallic region at intermediate $\Delta$'s.
The reason is that
there are sites where the random energies compensate a dynamically
generated on-site energy induced by the interaction with phonons. This
interaction produces
a checkerboard pattern of charge, leading to sites that are either
charge localized $n$=$1$ (negative effective energy)
or empty $n$=$0$ (positive effective energy). The compensating random energy
induces some of the sites to prefer charge $n$$\sim$$0.5$, as in a homogeneous metal.
It is through these sites that charge can
move from one side of the cluster to the other.
Compatible with this picture -- where only a fraction of the sites
have $n$ approximately $0.5$ -- the metallic state is found to have
poor-metallic behavior,
and in the histogram of conductances for different realizations of
quenched disorder many of them have insulating character.
The mechanism here discussed to understand the filling of gaps by disorder
appears general, and it should be valid beyond manganites.
In fact, there are experimental results and theoretical investigations
 that have reported similar effects
in the context of disordered insulators (see
Ref.~\onlinecite{pettifor} and references therein). In these
related investigations random configurations of disorder were
found to provide conducting resonant electronic channels for
tunneling in tunnel junctions, an explanation similar to that
provided in the present investigations. In addition, it is 
known that several materials share phenomenological aspects with
the manganites,\cite{clustered} and the investigations of these
interesting effects may have consequences in a variety of 
research areas.

Is this mechanism active in the real manganites? More work is
needed to investigate this issue. Reasons for concern are the
dependence of the results with the form of disorder, and the
irrelevance of the competing FM state in the process. Disorder in
Mn-oxides is believed to arise from lattice distortions induced by
chemical doping with ions that are either much larger or smaller than
those involved in the parent undoped compound. From this perspective
disorder in the hopping amplitudes would be realistic.
On the other hand,
the ionic dopants have often a different valence than the ion they
replaced and that could be represented by on-site energies. 
Additional effort is needed to relate this simple insulator-metal mechanism
to those proposed in early investigations and to experiments. 
However, even if the
immediate relevance of the idea to Mn-oxides needs further study,
the effort of Refs.~\onlinecite{aliaga,motome} and ours confirms the key
role that disorder plays to render metallic an insulating system,
a somewhat counter-intuitive idea. This adds further evidence that
the most basic aspects of the original scenario by Burgy {\it et
al.} \cite{burgy} are essentially correct, namely disorder is
needed at least as a triggering effect to induce insulator to
metal transitions in models for manganites in the vicinity of
regions where FM-AF phase competition occurs. Moreover, the present
effort has shown that the metallic state induced by disorder
is ``poor'' due to its inhomogeneous characteristics. All these results 
provide further support for the mixed-phase-based  percolative
explanation of the remarkable DC transport behavior of the
CMR manganites.

\vspace{7 mm}

\section{acknowledgments}

The authors are very thankful to Y. Motome, N. Furukawa, and N.
Nagaosa for sharing their ideas about Ref.~\onlinecite{motome}
with us, and for comments on the manuscript. The authors also
acknowledge the help of J. A. Verg\'es in the study of the
conductance. The subroutines used in this context were kindly
provided by him. The authors are supported by the NSF grants
DMR-0122523, DMR-0312333, and DMR-0303348. 
Additional funds have been provided by Martech (FSU).

\vspace{7 mm}

\section{Appendix: Issues related with Anderson localization}

It is clear in our figures for the
conductivity in 2D and 3D, corresponding to the simplified ``toy'' 
model, that $\sigma$ (or the conductance) reaches its maximum value 
at some intermediate disorder strength $\Delta$$\sim$$0.8$-$0.9$ (see
Figs.~\ref{fig:norcon} and \ref{fig:3dcon}).
However, both in 2D and 3D the conductance maximum value decreases
slowly as the lattice size increases. As discussed in the text,
this prevents a clear statement regarding the stability of a 
metallic phase in the bulk limit
by the mechanism discussed in this paper, although the
reasons for the metallicity tendencies were properly clarified.
The main concern about finding a metal in the bulk is related with
issues of Anderson localization, particularly in 2D, 
as discussed in this appendix.

In classical transport theory, which relies on weak
scattering, the conductance $G$ of a {\it{d}}-dimensional
hypercube of volume $L^d$ is related with the conductivity $\sigma$
through $G$=$\sigma L^{d-2}$, where $L$ is much bigger than the
mean free path, $l$. However,
Abrahams {\it et. al.} have shown\cite{abrahams,lee} -- based on
renormalization group ideas and perturbation theory -- that there is
no mobility edge in 2D, hence the conductance should
vanish as $L$$\rightarrow$$\infty$, where quantum interference
leads to Anderson localization. According to this
picture, the correct scaling of the conductivity in one, two, and
three dimensions is given by\cite{lee}:
$\sigma_{\rm 1D}(L)$=$\sigma_0 - \frac{2e^2}{h}(L-l)$, 
$\sigma_{\rm 2D}(L)$=$\sigma_0 - \frac{2e^2}{h\pi}\ln(\frac{L}{l})$,
and $\sigma_{\rm 3D}(L)$=$\sigma_0 -
\frac{2e^2}{h\pi^2}(\frac{1}{l}-\frac{1}{L})$,
respectively, where $\sigma_0=ne^2\tau/m$ is the Drude
conductivity at scale $l$ and $\tau=l/\hbar k_{\rm F}$ is the
relaxation time. In 3D, it seems possible to have a non-zero
conductance for macroscopic lattice sizes, but not in a lower
dimension.

To compare these formulas for localization with our
simulations, we proceed as follows. In 2D and 3D, the conductivity
attains its maximum value at the disorder strength $\Delta$=$0.8$ and
$\Delta$=$0.9$, respectively. We define the difference between
conductivities for two lattices of sizes $L_1$ and $L_2$ as:
\begin{eqnarray}
\Delta\sigma_{\rm 2D}^{\rm toy}(L_1,L_2)&=&\sigma_{\rm
2D}^{\Delta=0.8}(L_2)-\sigma_{\rm 2D}^{\Delta=0.8}(L_1),
\\
\Delta\sigma_{\rm 3D}^{\rm toy}(L_1,L_2)&=&\sigma_{\rm
3D}^{\Delta=0.9}(L_2)-\sigma_{\rm 3D}^{\Delta=0.9}(L_1),
\end{eqnarray}
where ``toy'' indicates that these are the numerical results for
the toy model. The corresponding ``localization'' formulas are
given by:
\begin{eqnarray}
\Delta\sigma_{\rm 2D}^{\rm
loc}(L_1,L2)&=&\frac{2}{\pi}\ln(\frac{L_1}{L_2}),
\\
\Delta\sigma_{\rm 3D}^{\rm
loc}(L_1,L2)&=&\frac{2}{\pi^2}(\frac{1}{L_2}-\frac{1}{L_1}).
\end{eqnarray}
In all of the above, the
units for conductivity are ``$e^2/h$''. The comparison between
these formulas is summarized in Table~\ref{tab:comp2D}, for the 2D case.

\begin{table}[h]
\caption{\label{tab:comp2D}Differences in
conductivities between pairs of lattices in 2D, using 
data from both the toy model and localization formulas.}
\begin{ruledtabular}
\begin{tabular}{cccc}
$L_2$&$L_1$&$\Delta\sigma_{\rm 2D}^{\rm
toy}(L_1,L_2)$&$\Delta\sigma_{\rm 2D}^{\rm
loc}(L_1,L2)$\\
\hline
$24$&$20$&$-0.1167$&$-0.1161$\\
$24$&$16$&$-0.1844$&$-0.2581$\\
$24$&$12$&$-0.3735$&$-0.4413$\\
$20$&$16$&$-0.0677$&$-0.1421$\\
$20$&$12$&$-0.2568$&$-0.3252$\\
$16$&$12$&$-0.1891$&$-0.1831$\\
\end{tabular}
\end{ruledtabular}
\end{table}

The similarity of the numbers leads to the 
conclusion that the numerical results
for the 2D toy model in the ``metallic regime'' could
still lead to an Anderson insulator as the lattice size grows.
However, obviously the mechanism discussed in this paper is
mainly to be applied to 3D manganites (or bilayered systems).
It is known that Anderson localization
is not so severe in 3D, giving confidence that the state
stabilized at intermediate $\Delta$'s is indeed a metal in the
bulk limit.


\begin{thebibliography}{99}
\bibitem{comment-density}
 The use of, e.g.,
$\langle n_{\mathbf{i}}\rangle \langle n_{\mathbf{i+x}} \rangle$ instead
of $\langle n_{\mathbf{i}} n_{\mathbf{i+x}} \rangle$ is merely for technical
convenience, and at low temperatures it does not affect our main conclusions.
\end{thebibliography}
\end{document}